\documentclass{rmf-d}
\usepackage{nopageno,rmfbib,multicol,times,epsf,amsmath,amssymb,cite}
\usepackage[latin1]{inputenc}
\usepackage[]{caption2}
\usepackage{graphics}
\usepackage[demo]{graphicx}
\usepackage{hyperref}
\usepackage{lipsum}
\usepackage{wrapfig, blindtext}
\usepackage{comment}
\usepackage{tabularx}
\usepackage{float}
\usepackage{slashed}
\def\rmfcornisa{Research OR Education of Physics \hfill\rmf\ {\bf xx} (*x*) xxx--xxx
\hfill MES ---A\~NO 2022}

\def\rmfcintilla{{\it Rev.\ Mex.\ Fis.\/} {\bf ??} (*?*) (????) ???--???}
\clearpage \rmfcaptionstyle 
\begin{document}
\title{  The decay of $N^*$(1895) to light hyperon resonances
}
\author{K. P. Khemchandani$^1$, A. Mart\'inez Torres$^2$, Sang-Ho Kim$^3$, Seung-il Nam$^{3,4}$, H. Nagahiro$^5$, A. Hosaka$^6$}
\address{ 
$^1$Universidade Federal de S\~ao Paulo, C.P. 01302-907, S\~ao Paulo, Brazil.\\
$^2$Universidade de Sao Paulo, Instituto de Fisica, C.P. 05389-970, Sao Paulo, Brazil.\\
$^3$Department of Physics, Pukyong National University (PKNU), Busan 48513, Korea.\\
$^4$Asia Pacific Center for Theoretical Physics (APCTP), Pohang 37673, Korea.\\
$^5$Department of Physics, Nara Women's University, Nara 630-8506, Japan.\\
$^6$Research Center for Nuclear Physics (RCNP), Osaka University, Ibaraki, Osaka, 567-0047, Japan.  }
\maketitle
\begin{abstract}
\vspace{1em}In this talk I review the findings of our recent works where we have studied the decay of $N^*(1895)$ and the implications of such properties on the photoproduction of light hyperons. I discuss that meson-baryon interactions play an essential role in describing the nature of $N^*(1895)$ and report the details of our investigation of its  decays to different meson-baryon systems and to final states involving $\Lambda(1405)$ and a proposed $\Sigma(1400)$. We find  that the width of $N^*(1895)$ gets  important contributions from the decay to light hyperon resonances. Such an information can be used to look for alternative processes  to study $N^*(1895)$ in experimental data.
\vspace{1em} 
\end{abstract}
\keys{Nucleon resonances, hyperons, decay process}
\begin{multicols}{2}

\section{Introduction}
Determining the properties of nuclear resonances from experimental data gets challenging very quickly as one goes to higher energies in the spectrum. One of the frequently faced difficulties is that the states become wider and couple to common meson-baryon channels. An example of the consequence of such difficulties is the fact that all the $1/2^-$ structures with mass higher than 1800 MeV are clubbed together under the heading of $N^*(1895)$ in the particle data book~\cite{pdg}. In fact, nucleon states around 1890 MeV have been the focus of a debate during the last decade since different descriptions have been brought forward  for the peak present around 1900 MeV in the $\gamma p \to K^+ \Lambda$ total cross sections. As can be seen from Table~\ref{Table:I}, the particle data group (PDG)~\cite{pdg} catalogues several nucleon states in this energy region.  
\begin{table}[H]
\caption{Cluster of nucleon states present around 1890 MeV (as taken from the Ref.~\cite{pdg}). }\label{Table:I}
\begin{center}
\begin{tabular}{cccc}
\hline\\
State& Spin-parity  & Mass &Width \\
& $\left(J^\pi\right)$ &  (MeV)& (MeV)\\
\hline\hline\\
$N^*(1875)$&$3/2^-$& 1850 to 1950&100 to 220\\
$N^*(1880)$&$1/2^+$&1820 to 1900 &180 to 280\\
$N^*(1895)$&$1/2^-$&1890 to 1930 &80 to 140\\
$N^*(1900)$&$3/2^+$&1900 to 1940& 100 to 200\\
\hline
\end{tabular}
\end{center}
\end{table}
In fact $N^*(1895)$ was listed as  $N^*(2090)$ ($J^\pi=1/2^-$) in the compilations earlier than  2012 of the PDG and instead of the current $S_{11}$  state a proposal for the existence of a $3/2^+$ $N^*(1895)$ was discussed   in Ref.~\cite{Mart:1999ed} in order to describe the experimental data on $\gamma p \to K^+ \Lambda$. Such  a state was predicted by the quark model of Ref.~\cite{Capstick:1993kb}.  Later partial wave investigations~\cite{Anisovich:2011fc,Mart:2019mtq,Mart:2012fa}, however,  concluded that the peak corresponds to the presence of a $3/2^+$ $N^*(1900)$ state. An alternative  description for the same peak in the data is provided  in Ref.~\cite{MartinezTorres:2009cw},  indicating the presence of a $1/2^+$ nucleon resonance with mass around 1900 MeV.  The complication arises from the fact that data on reactions producing $K\Lambda$, $K\Sigma$ are mainly used  in partial wave analyses to study $N^*$ states in this energy region  and most states couple to both these channels.  In such a situation it is important to consider other processes to better distinguish different nucleon resonances. Indeed, as a step forward in this direction, the $K^*(892)\Lambda$  final state has been considered in Ref.~\cite{Anisovich:2017rpe}. Our study of decay properties of $N^*(1895)$ is an effort to suggest alternative sources of information on the nucleon resonances around 1900 MeV.

We must mention here the additional and the initial motivation of our study of decay properties of  $N^*(1895)$. This state lies very close to the thresholds of  $K\Lambda(1405)$ and $K\Sigma(1400)$, where $\Sigma(1400)$ is a $J^\pi = 1/2^-$ state whose existence was proposed in a recent study~\cite{Khemchandani:2018amu} carried out by some of the co-authors of this manuscript. It must be added that the work in Ref.~\cite{Khemchandani:2018amu} is certainly not the first one indicating the existence of a $\Sigma$ state near 1400 MeV. The need for the existence of such an isovector resonance has been discussed when describing different experimental data in Refs.~\cite{Oller:2000fj,Guo,Wu:2009tu,Wu:2009nw,Gao:2010hy,Xie:2014zga,Xie:2017xwx}. In Ref.~\cite{Roca:2013cca}, though, a strong cusp around the $\bar K N$ threshold, and not a state, is found when analysing the data on $\Lambda(1405)$ photoproduction. In such a scenario it can be useful to encourage experimental investigation of photoproduction of an isovector partner of $\Lambda(1405)$ and to do the same we must provide an estimate of the cross sections of the process.  It was shown in Ref.~\cite{Kim:2017nxg} that the exchange of nucleon resonances in the s-channel play an important role in describing  the data on $\gamma p \to K^+ \Lambda(1405)$. Thus, in order to  evaluate  the photoproduction cross sections for $\Sigma(1400)$ we must study the process  $N^*(1895) \to K\Sigma(1400)$ .   It can be useful to determine the decay width for $N^*(1895)\to K\Lambda(1405)$ as well and study its contribution to the $\Lambda(1405)$ photoproduction. We can obtain the decay rate for  these processes reliably since we have earlier studied pseudoscalar/vector-baryon coupled channel scattering~\cite{Khemchandani:2013nma} and found $N^*(1895)$ to arise from the underlying dynamics.  Consequently,  the couplings of different meson-baryon channels to  $N^*(1895)$ are known from our previous work.  Our findings can also be useful to study  the process  $\pi N \to  K^* \pi \Sigma$, which is expected to be investigated at J-PARC~\cite{Noumi:2017sdz}.

\section{Formalism}
\begin{figure}[H]
    \centering
    \includegraphics[width=.9\linewidth]{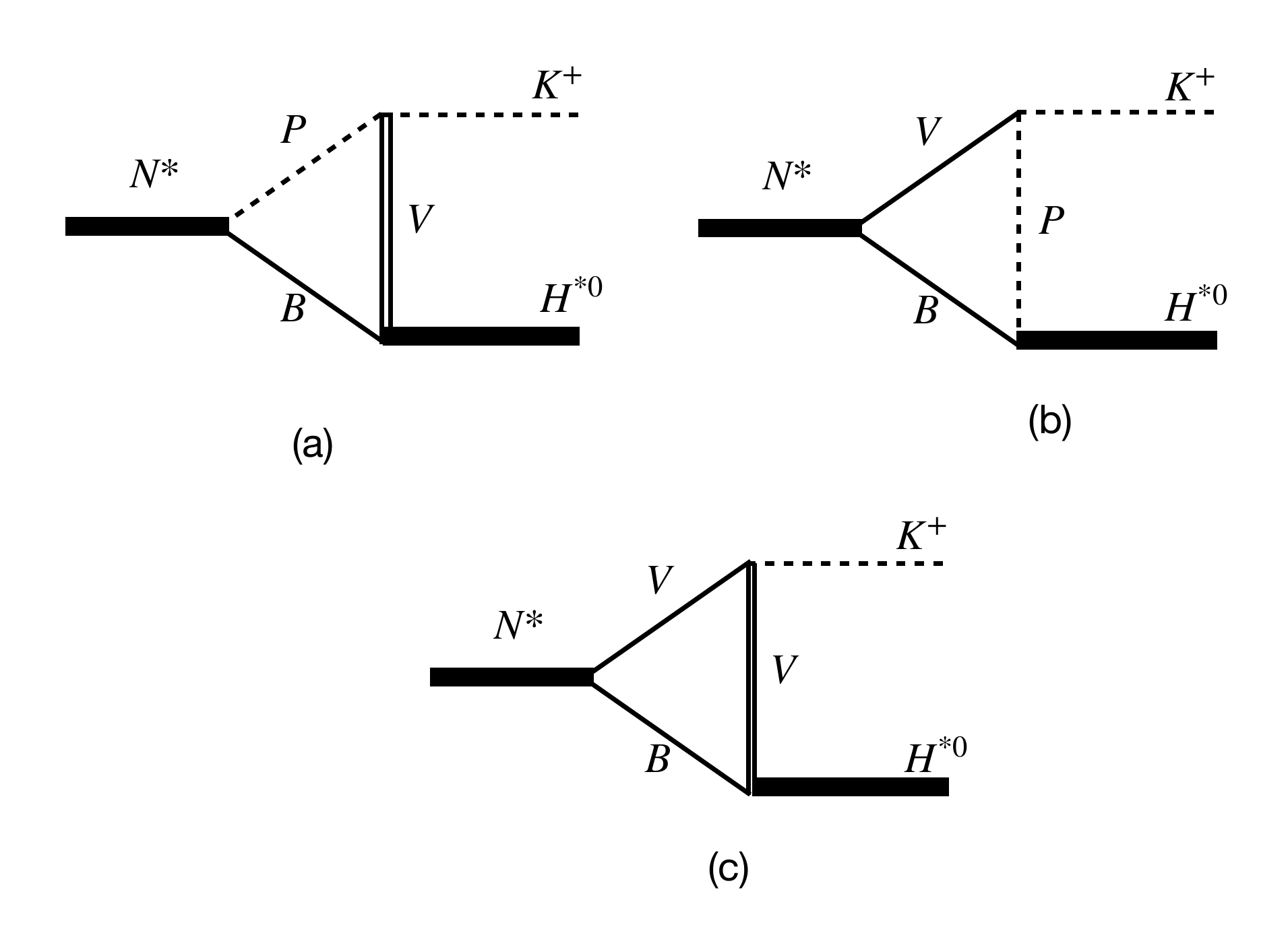}  
    \caption{Different diagrams contributing to the decay of $N^*(1895)$ to $K\Lambda(1405)$ and $K\Sigma(1400)$.}
\label{diagrams}
\end{figure}

It was shown in our former work~\cite{Khemchandani:2013nma}  that  $N^*(1895)$ couples strongly to pseudoscalar/vector-baryon  channels and is associated with two poles in the complex plane. Our findings of the relevance of  hadron dynamics in describing $N^*(1895)$ should come as no surprise to the reader. We must recall that the properties of $N^*(1895)$ do not agree with those obtained for a third $1/2^-$ nucleon, after $N^*(1535)$ and $N^*(1650)$, within different quark models~\cite{Isgur:1978xj,Bijker:1994yr,Hosaka:1997kh,Takayama:1999kc}. Its mass, for example, is estimated to be  $>$ 2100 MeV in Refs.~\cite{Hosaka:1997kh,Takayama:1999kc} when considering harmonic oscillator potential.  In Ref.~\cite{Khemchandani:2013nma}, on the other hand, we obtained the mass and width of  $N^*(1895)$ in good agreement with the values of PDG~\cite{pdg}. Besides, different amplitudes in Ref.~\cite{Khemchandani:2013nma} reproduce the available  experimental data on total cross sections for processes, like, $\pi^- p \to \eta n$, $K^0 \Lambda$, as well as the $\pi N$ scattering amplitudes, in isospin 1/2 and  3/2, known from the partial wave analysis of relevant data.  It is well known that meson-baryon dynamics play an important role in describing $\Lambda(1405)$ as well~(for a recent review see Ref.~\cite{Mai:2020ltx}  and to see some well cited articles see Refs.~\cite{osetramos,Oller:2000fj,Jido:2003cb,Hyodo:2011ur,Mai:2014xna} ). Indeed  a similar nature  was  proposed  for $\Sigma(1400)$ also in Ref.~\cite{Khemchandani:2013nma}. From all these former findings, we can describe the decay of $N^*(1895)$ to light hyperons  through the diagrams shown in Fig.~\ref{diagrams}. 

We would like to emphasize here that the couplings for all the vertices in Fig.~\ref{diagrams} are known from our previous works~\cite{Khemchandani:2013nma,Khemchandani:2018amu} which describe relevant experimental data. In the study of hyperon resonances of Ref.~\cite{Khemchandani:2018amu}, for example, we reproduce experimental data on the total cross sections of $K^- p \to K^- p$, $\bar K^0 n$, $\eta \Lambda$, $\pi^0 \Lambda$, $\pi^0 \Sigma^0$, $\pi^\pm \Sigma^\mp$ and the data on the energy level shift and width of the $1s$ state of the kaonic hydrogen.  

Let us now discuss the Lagrangians needed to write the amplitudes for the diagrams in Fig.~\ref{diagrams}. The vertices involving the nucleon/hyperon resonances are written as
\begin{align}
&\mathcal{L}_{N^*PB}=i g_{PBN^*} \bar B N^* P^\dagger , \nonumber\\
&\mathcal{L}_{N^*VB}=-i\frac{g_{VBN^*}}{\sqrt{3}} \bar B \gamma_5 \gamma_\mu N^* V^{\mu^\dagger} , \nonumber\\
&\mathcal{L}_{PBH^*}=g_{PBH^*} P \bar H^* B , \nonumber\\
&\mathcal{L}_{VBH^*}=i\frac{g_{VBH^*}}{\sqrt{3}} V^{\mu} \bar H^*  \gamma_\mu \gamma_5 B,\label{lageff}
\end{align}
and the remaining ones as
\begin{align}
&\mathcal{L}_{PPV}=-i g_{PPV}\langle V^\mu \left[ P, \partial_\mu P \right] \rangle, \label{vpp} \\
&\mathcal{L}_{VVP}=\frac{g_{VVP}}{\sqrt{2}}\epsilon^{\mu \nu \alpha \beta} \langle \partial_\mu V^\nu \partial_\alpha V_\beta P \rangle. \label{vvp}
\end{align}
The field $H^*$ in  the above equations represents $\Sigma(1400)$ or $\Lambda(1405)$, and the factor $\sqrt{3}$ in the Lagrangians, $\mathcal{L}_{N^*VB}$ and $\mathcal{L}_{VBH^*}$, appears due to the fact that the spin-projected amplitudes were parameterized as Breit-Wigner in Refs.~\cite{Khemchandani:2013nma,Khemchandani:2018amu} when calculating the meson-baryon-resonance couplings.   

Using the mentioned Lagrangians we obtain the following amplitudes for the three diagrams in Fig.~\ref{diagrams}
\begin{align}
t_a = &i  \sum_j g_{VBH^*\!, j}~ g_{PBN^*\!, j} ~g_{PPV} ~C_j ~\bar u_{H^*}\left(p\right)\gamma_\nu \gamma_5 \nonumber
\\&\int\frac{d^4q}{(2\pi)^4}\Biggl\{ \frac{\left(\slashed{P}-\slashed{k}+\slashed{q}+m_{Bj}\right)}{\left(P-k+q\right)^2-m^2_{Bj}+i\epsilon}\Biggr.\nonumber\\
&\times\left.\frac{\left(-g^{\nu\mu}+\dfrac{q^\nu q^\mu}{m^2_{Vj}}\right)}{q^2-m^2_{Vj}+i\epsilon}\frac{\left(2k -q\right)_\mu}{\left(k-q\right)^2-m^2_{Pj}+i\epsilon}\right\} u_{N^*}\left(P\right),\label{ta1}
\end{align}
\end{multicols}
\begin{align}
t_b=&- \sum_j g_{PBH^*\!, j}~ g_{VBN^*\!, j} ~g_{PPV} ~D_j ~\bar u_{H^*}\left(p\right) 
\int\frac{d^4q}{(2\pi)^4}\Biggl\{ \frac{\left(\slashed{P}-\slashed{k}+\slashed{q}+m_{Bj}\right)}{\left(P-k+q\right)^2-m^2_{Bj}+i\epsilon} \gamma_5\gamma_\nu\Biggr.\nonumber\\
&\times\left.\frac{\left(-g^{\nu\mu}+\dfrac{\left(k-q\right)^\nu \left(k-q\right)^\mu}{m^2_{vj}}\right)}{\left(k-q\right)^2-m^2_{vj}+i\epsilon}\frac{\left(k +q\right)_\mu}{q^2-m^2_{pj}+i\epsilon}\right\} u_{N^*}\left(P\right),\label{tb1}
\end{align}
and 
\begin{align}
&t_c=i \sum_j g_{VBH^*\!, j}~ g_{VBN^*\!, j} \frac{g_{VVP}}{\sqrt{2}} ~F_j ~\bar u_{H^*}\left(p\right)\int\frac{d^4q}{(2\pi)^4}\left\{\epsilon^{\lambda\nu\alpha\beta}\frac{\left(-g^\sigma_\beta+\dfrac{q^\sigma q_\beta}{m^2_{vj_1}}\right)}{\left(k-q\right)^2-m^2_{vj_1}+i\epsilon} \gamma_\sigma\gamma_5\right.\nonumber\\
&\times\frac{\left(\slashed{P}-\slashed{k}+\slashed{q}+m_{Bj}\right)}{\left(P-k+q\right)^2-m^2_{Bj}+i\epsilon} \gamma_5\gamma_\mu \left.\frac{\left(-g^{\mu}_\nu+\dfrac{\left(k-q\right)^\mu \left(k-q\right)_\nu}{m^2_{vj_2}}\right)}{q^2-m^2_{vj_2}+i\epsilon}\left(k-q\right)_\lambda q^\alpha\right\} u_{N^*}\left(P\right),\label{tc1}
\end{align}
\begin{multicols}{2}
where the constants $D_j $ and $F_j$ come from the trace in Eq.~(\ref{vpp}) whose values are given in Ref.~\cite{Khemchandani:2020exc}. The integration on $q^0$ in all the amplitudes is done analytically and the results are given in Ref.~\cite{Khemchandani:2020exc}. The integration on three-momentum is done numerically. We must add here that the final state needs to be projected on p-wave (for more details see Ref.~\cite{Khemchandani:2020exc}). 

\begin{figure}[H]
    \centering
    \includegraphics[width=.9\linewidth]{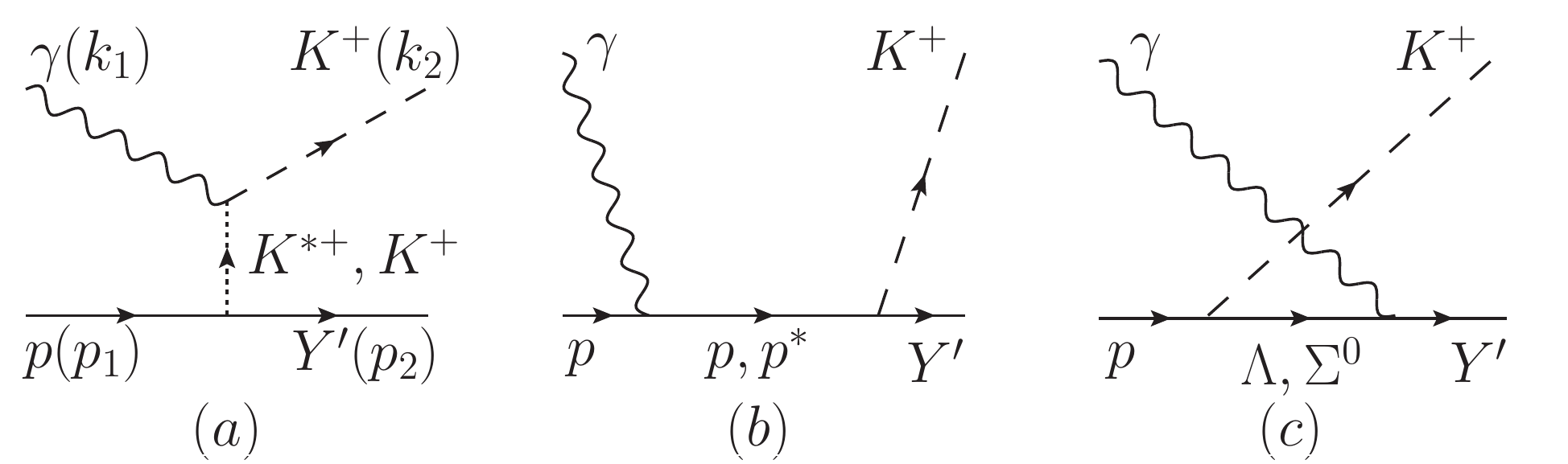}  
    \caption{Diagrams required to study the photoproduction of $\Lambda(1405)$ and $\Sigma(1400)$.}
\label{diagramphoto}
\end{figure}

\begin{figure}[H]
    \centering
    \includegraphics[width=.6\linewidth]{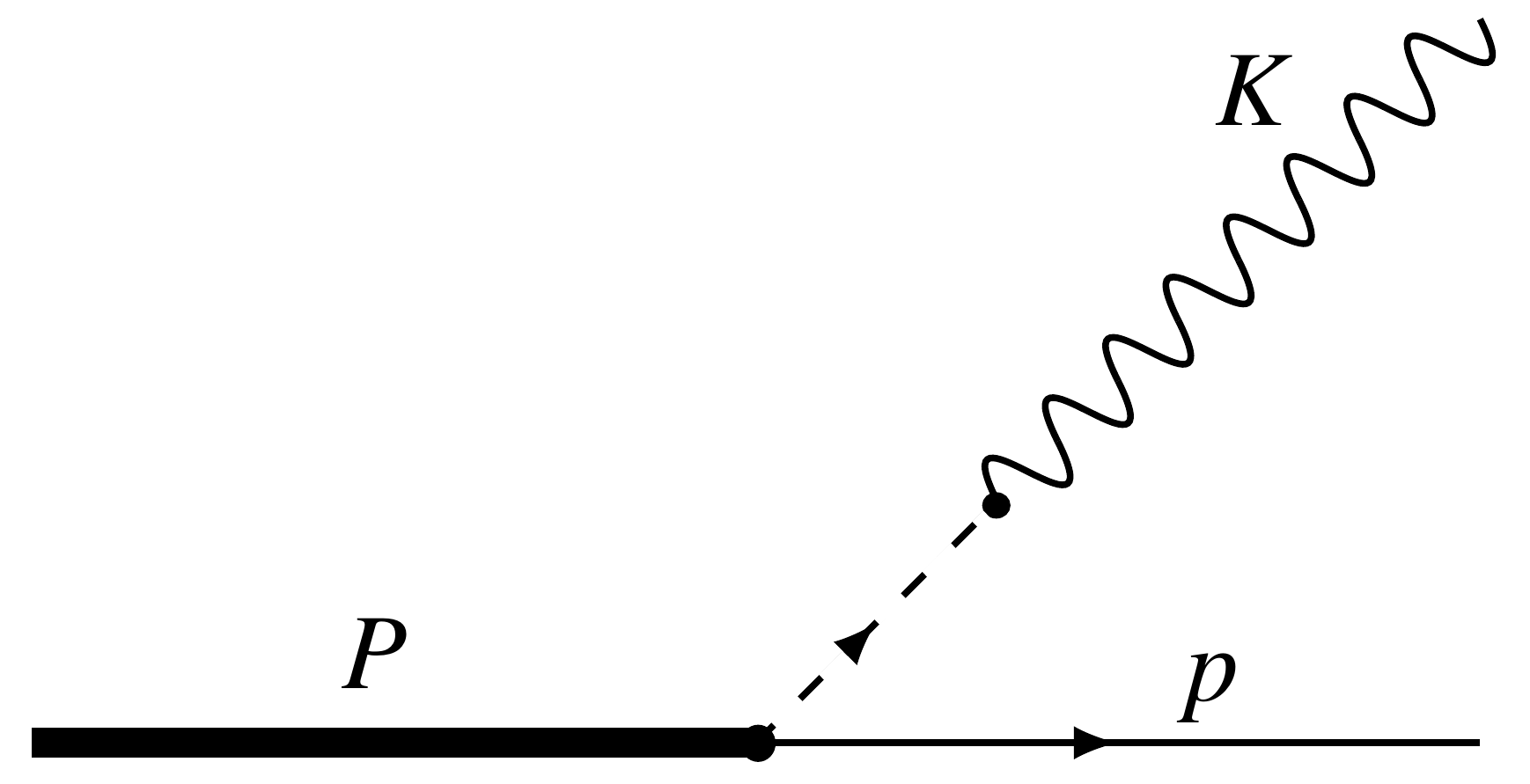}  
    \caption{Radiative decay of a resonance through the vector meson dominance. The dashed line in the figure represents a vector meson.}
\label{vmd}
\end{figure}

With the couplings available from Refs.~\cite{Khemchandani:2013nma,Khemchandani:2018amu} we can also calculate the radiative decay widths of the nucleon/hyperon resonances. Such widths can be useful in the calculation of the diagrams contributing to the photoproduction of $\Lambda(1405)/\Sigma(1400)$ (as shown in Fig.~\ref{diagramphoto}).
We calculate the radiative width considering the vector meson dominance mechanism (see Fig.~\ref{vmd}). For this purpose, we consider the Lagrangian~\cite{Roca:2003uk}
\begin{align}
\mathcal{L}_{V\gamma} =-\frac{e F_V}{2} \lambda_{V\gamma} V_{\mu \nu} A^{\mu \nu},\label{vgamma}
\end{align}
where $F_V$ is the decay constant for vector mesons, $A^{\mu \nu}=\partial^\mu A^\nu-\partial^\nu A^\mu$, $V_{\mu \nu}$ is a tensor field related to $\rho^0$, $\omega$, $\phi$, with $\lambda_ {V\gamma} = 1,~\frac{1}{3},~-\frac{\sqrt{2}}{3}$, respectively, and 
\begin{align}
V^{\mu \nu}= \frac{1}{M_V}\left(\partial^\mu V^\nu-\partial^\nu V^\mu\right).
\end{align}
Consequently, we obtain the amplitude for the $B^* \to B \gamma$ process  as
\begin{align}
t_{B^*\to B \gamma} = \frac{2 e F_V \tilde g_{VBB^*} \lambda_{V\gamma}}{ M_V^2} \bar B \gamma_5 \slashed \epsilon \slashed K B^*,
\end{align}
with $\epsilon$ denoting the polarization vector for the photon and plug it in the following equation to determine the width,
 \begin{align}
 \Gamma_{B^* \to B \gamma} &= \frac{1}{32 \pi^2}\frac{|~\vec K~|\left( 4 M_{B^*} M_{B}\right)}{M_{B^*}^2}\frac{1}{2 S_{B^*} +1}\nonumber\\&\times \int d\Omega \sum_{m_{B^*}, m_{B}, m_\gamma} |\mathcal{M}_{B^* \to B \gamma}|^2,\label{eq:width2}
 \end{align}

\section{Results and discussion}
\begin{figure}[H]
    \centering
    \includegraphics[width=.8\linewidth]{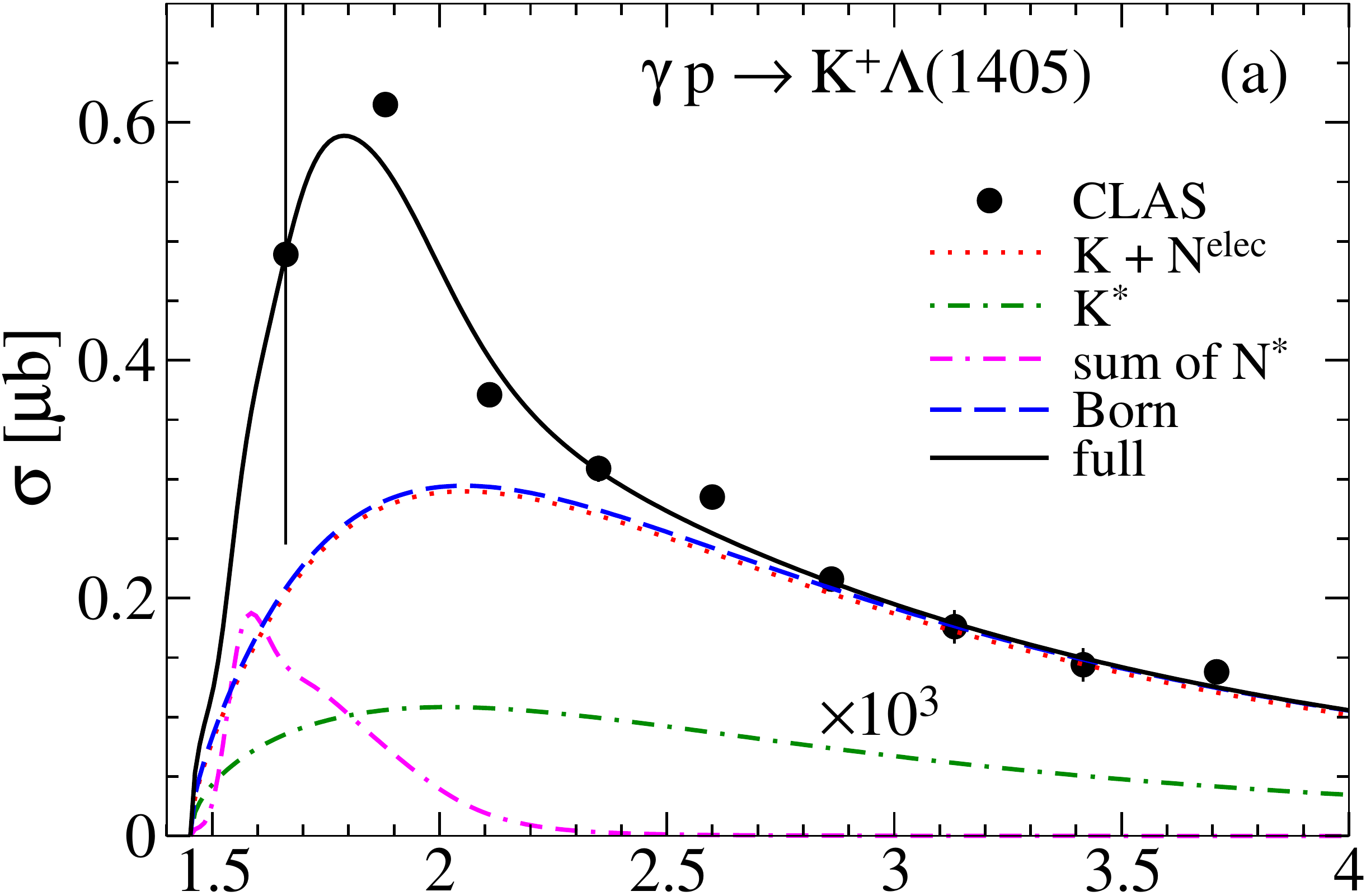}  
    \caption{Total cross section for $\gamma p \to K^+ \Lambda(1405)$  
as a function of the beam energy, $E_\gamma$.
The (magenta) dot-double-dashed curve depicts, among other $N^*$'s, the
contribution from the $N^*(1895)$ exchange, which happens to be the dominant.
The data are taken from  Ref.~\cite{Moriya:2013hwg}.
}
\label{XnLambda}
\end{figure}

\begin{figure}[H]
    \centering
    \includegraphics[width=.8\linewidth]{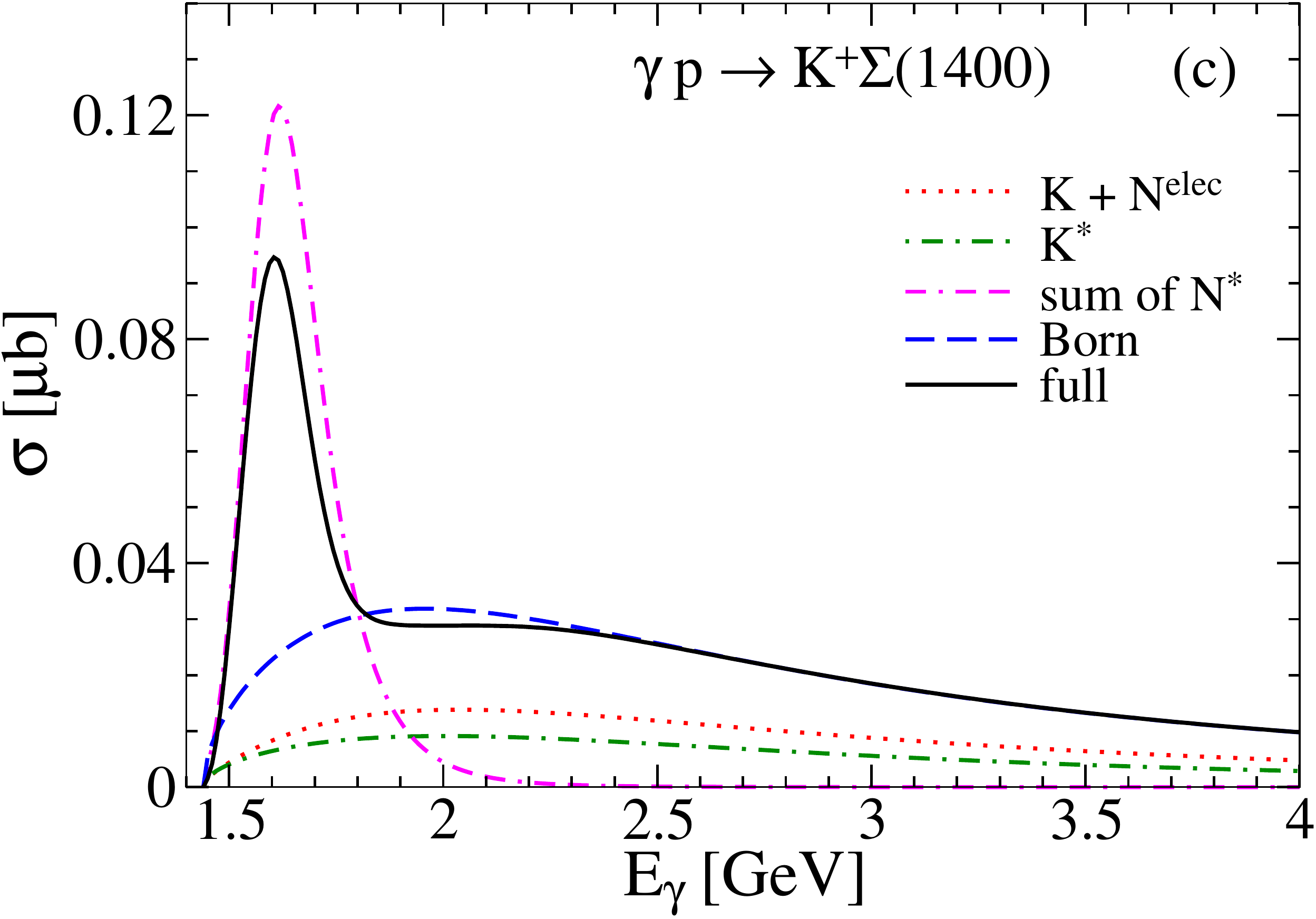}  
    \caption{Total cross section for $\gamma p \to K^+ \Sigma(1400)$ 
as a function of the beam energy, $E_\gamma$.
Here  the (magenta) dot-double-dashed curve depicts the contribution from the s-channel diagram with the $N^*(1895)$ exchange.}
\label{XnSigma}
\end{figure}

Let us first discuss the results on the decay width of $N^*(1895)$. The amplitudes of the diagrams given in Fig.~\ref{diagrams} are summed coherently to calculate the decay to the final state with hyperon resonances through 
  \begin{align}
 \Gamma_{N^* \to K H^*} &= \frac{1}{32 \pi^2}\frac{|~\vec p~|\left( 4 M_{H^*} M_{N^*}\right)}{M_{N^*}^2}\frac{1}{2 S_{N^*} +1}\nonumber\\&\times \int d\Omega \sum_{m_{N^*}, m_{H^*}} |t_{N^* \to K H^*}|^2,\label{eq:width}
 \end{align}
where $H^*$ denotes the hyperon resonance, $\Sigma^*$ or $\Lambda^*$.  We recall at this point that both $N^*(1895)$ and $\Lambda(1405)$ are related to two poles in our model. It is useful to present the related poles for further discussions and we do so in Table~\ref{poles}.
\begin{table}[H]
\caption{The poles related to $N^*(1895)$, $\Lambda(1405)$ and $\Sigma(1400)$ as obtained in Refs.~\cite{Khemchandani:2013nma,Khemchandani:2018amu}. Notice that two poles are associated with $N^*(1895)$ and $\Lambda(1405)$.}\label{poles}
\centering
\begin{tabular}{ccc}
\hline\hline
State&  \multicolumn{2}{c}{Pole position (MeV)}\\
& \multicolumn{2}{c}{ $E- i \Gamma/2$}\\\hline
$N^*(1895)$ & $1801-i96\quad$&$1912-i54$\\
$\Lambda(1405)$&$1385-i124\quad$&$1426-i15$\\
$\Sigma(1400)$&\multicolumn{2}{c}{$1399 - i 36$}\\\hline\hline
\end{tabular}
\end{table}
For the sake of clarity we refer to the lower (upper) poles of $N^*$ and $\Lambda^*$ as $N^*_1(N^*_2)$ and $\Lambda^*_1(\Lambda^*_2)$, respectively.  It can be noticed from Table~\ref{poles} that  the  $N^*$ mass (the central value of the poles) lies below the threshold of some  decay channels. We must, however, remember that each  of  the $N^*$ poles has a finite width, which will contribute to the decay rate. We have considered this fact by convoluting the decay width over the varying mass of $N^*_i$. The results, thus,  obtained are shown in Table~\ref{Table1} in terms of the branching ratios. It can be seen that our results on decay widths to different pseudoscalar/vector-baryon channels are in good agreement with experimental data. Further, the decay widths to light hyperons are comparable with those to meson- baryon channels. Considering the fact that  the widths of the two poles are overlapping in case of $N^*(1895)$ as well as $\Lambda(1405)$, and it can be difficult to distinguish such poles, it can be useful to provide the results where we consider interference of the amplitudes related to the different poles. In such a case we use an average mass of approximately 1895 MeV and a width of approximately 120~MeV for $N^*(1895)$ in the phase space and get
\begin{align}
\Gamma_{N^{*}(1895)\to K\Sigma(1400)}=\left(18.9 \pm 1.5\right)  \text{ MeV},\label{GS}\\
\Gamma_{N^{*}(1895)\to K \Lambda(1405)}= \left(8.3 \pm 1.3\right) \text{ MeV}.\label{GL}
\end{align}
These values of decay widths to $KH^*$  indicate that processes, like, the photoproduction of $K\Lambda(1405)$ can be a useful source of information on the properties of $N^*(1895)$, in addition to the processes with $K\Lambda$, $K\Sigma$ final states considered in usual partial wave analysis.  We also provide the results on radiative decay widths considering interference of the amplitudes related to the two poles of $N^*(1895)$ and $\Lambda(1405)$ in Table~\ref{TAB3}. We can  compare the results with the information available from the experiments. The radiative decay width of $\Lambda(1405)\to\Lambda \gamma$ determined from the experimental data is known to be  $27\pm8$~KeV~\cite{pdg}. Our result obtained by considering the superposition of the two poles is in remarkable agreement with the experimental data. For $\Lambda(1405)\to\Sigma \gamma$, PDG~\cite{pdg} provides two possible values: $10\pm4$~KeV or $23\pm7$~KeV. Our results are closer to the former value. 
\begin{table}[H]
\caption{Branching fractions of $N^*(1895)$ to different decay channels. The subscripts $1$, $2$ on $N^*$ and on $\Lambda$ refer to the respective lower and upper mass poles (as shown in Table.~\ref{poles}). Here we show the central values of the results. For details on the estimation of the uncertainties in our results we refer the reader to Ref.~\cite{Khemchandani:2020exc}. }
\label{Table1}
\centering
\begin{tabular}{cccc}
\hline\hline\\
Decay channel& \multicolumn{2}{c}{Branching ratio $(\%)$} & Experimental  \\ 
&  $N_1^*$ & $N_2^*$&  data~\cite{pdg} \\ 
\hline\\
$K \Lambda^*_1$&$5.4$ &$1.9$&--\\
$K \Lambda^*_2$&$3.4$&$1.1$&--\\
$K \Sigma^*$    &$5.9$&$11.4$&--\\
$\pi N$             &9.4    &  10.8     &   2-18  \\
$\eta N$           &2.7    &   18.1    &   15-40\\
$K \Lambda$   &10.9  &  19.4     &   13-23 \\
$K \Sigma$      &0.7    &   26.0    &    6-20\\
$\rho N$           &5.6    &  3.5     &     $<$18\\
$\omega N$     &25.7  &   6.2    &       16-40\\
$\phi N$             &8.9   &   1.1     &      --\\
$K^* \Lambda$ &12.1 &  14.0     &      4-9 \\
$K^* \Sigma$    &6.1   &   0.3    &      -- \\
\hline\hline
\end{tabular}
\end{table}
In case of $N^*(1895)$, the branching ratio of the radiative decay is given as 0.01-0.06 $\%$ in Ref.~\cite{pdg}. In our case, the branching ratio for $N_1^*$ turns out to be 0.34$-$0.42$~\%$, while for $N^*_2$ is 0.11$-$0.13$~\%$. Our results for the second pole seem to be closer to the upper limit of the value listed in Ref.~\cite{pdg}. Coincidently,  the real and imaginary part  of this second pole are closer to the values associated with $N^*(1895)$ in Ref.~\cite{pdg}.
\begin{table}[H]
\caption{Radiative decay widths for $\Lambda(1405)$, $\Sigma(1400)$ and $N^*(1895)$. The underlined process means that an interference between the two poles related to the decaying hadron has been considered to obtain the decay width.}
\label{TAB3}
\centering
\begin{tabular}{cc}
\hline\hline
Decay process& Partial width (KeV) \\ 
\hline
$\Lambda(1405) \to \Lambda \gamma$&26.19 $\pm$ 6.93\\
$N^*(1895) \to p \gamma$ &650.70 $\pm$ 65.10\\
$\Sigma(1400) \to \Lambda \gamma$&49.97 $\pm$ 8.57\\
$\Sigma(1400) \to \Sigma \gamma$&94.51$ \pm$ 9.33\\
$\Lambda(1405) \to \Sigma \gamma$&2.50 $\pm$ 1.37\\
\hline\hline
\end{tabular}
\end{table}
We would now like to show  the results of the cross sections of photoproduction of $\Lambda(1405)$. The amplitudes for the process is obtained by considering the diagrams shown in Fig~\ref{diagramphoto}. The  extension to higher energies is done by considering a Regge approach by using $K$- and $K^*$-Reggeon
exchange in the t-channel. More details on the formalism can be found in Ref.~\cite{Kim:2021wov}. The results on the total cross sections are shown in Fig.~\ref{XnLambda} as a function of the beam energy. It can be seen that our results are in good agreement with the data from the  CLAS Collaboration~\cite{Moriya:2013hwg}.  The point we would like to highlight here is the fact that the $N^*$ contributions play an important role in describing the data in the low-energy region ($E_\gamma \leqslant 2.5$ GeV). Apart from $N^*(1895)$, other states are included in the s-channel exchange, following the work of Ref.~\cite{Kim:2017nxg}:
$N^*(2000,\,5/2^+)$ $N^*(2100,\,1/2^+)$, 
$N^*(2030,1/2^-)$, $N^*(2055,\,3/2^-)$, and $N^*(2095,\,3/2^-)$. However, the contributions from the states other than $N^*(1895)$ are found to be small. Detailed comparisons can be found in Ref.~\cite{Kim:2021wov}.
We also predict the cross sections for  the process $\gamma p \to K^+ \Sigma(1400)$. The results are shown in Fig.~\ref{XnSigma}. The order of magnitude of the  cross sections  shown in  Fig.~\ref{XnSigma} are measurable in future. It should be pointed out that in this case too, the $N^*(1895)$ exchange in the s-channel provides important contributions to the cross sections near the threshold. We hope that our findings motivate future experimental investigations of the photoproduction of $\Sigma(1400)$. It should be mentioned that results on polarization observables are also shown and discussed in Ref.~\cite{Kim:2021wov}.
 \section*{Acknowledgements}
K.P.K and A.M.T gratefully acknowledge the  support from the Funda\c c\~ao de Amparo \`a Pesquisa do Estado de S\~ao Paulo (FAPESP), processos n${}^\circ$ 2019/17149-3 and 2019/16924-3, by the Conselho Nacional de Desenvolvimento Cient\'ifico e Tecnol\'ogico (CNPq), grants n${}^\circ$ 305526/2019-7 and 303945/2019-2. A.M.T also thanks the partial support from mobilidade Santander  (edital PRPG no 11/2019). H.N. is supported in part by Grants-in-Aid for Scientific Research (JP17K05443 (C)). AH is supported in part by Grants-in-Aid for Scientific Research (JP17K05441 (C)) and for Scientific Research on Innovative Areas (No. 18H05407). S.H.K. and  S.i.N thank National Research Foundation of Korea funded by the Ministry of Education, Science and Technology (MSIT) (NRF-2019R1C1C1005790 (S.H.K.), 2018R1A5A1025563, and 2019R1A2C1005697).

\end{multicols}
\medline
\begin{multicols}{2}

\end{multicols}
\end{document}